# An Exploration of a Discrete Rhombohedral Lattice

*of Possible Engineering or Physical Relevance*

**Jim McGovern**

Dublin Institute of Technology
School of Mechanical and Transport Engineering
Bolton Street, Dublin 1, Ireland
jim.mcgovern@dit.ie

A particular discrete rhombohedral lattice consisting of four symmetrically interlaced cuboctahedral point lattices is described that is interesting because of the high degree of symmetry it exhibits. The four constituent cuboctahedral lattices are denoted by four colours and the composite lattice is referred to as a 4-colour rhombohedral lattice. Each point of the 4-colour lattice can be referenced by an integer 4-tuple containing only the positive non-zero integers (the counting numbers). The relationship between the discrete rhombohedral lattice and a discrete Cartesian lattice is explained. Some interesting aspects of the lattice and of the counting-number 4-tuple coordinate system are pointed out.

## ■ Introduction

Symmetry underlies nature and has a major role to play in science [1], art, mathematics [2] and engineering. In two dimensions there are only three fully regular or symmetric point lattices and in three dimensions there is only one fully regular or symmetric point lattice [3]. These are the triangular tessellation, the square tessellation and the hexagonal tessellation in two dimensions and the cubic lattice in three. Here the expression *fully regular or symmetric* is used in the sense that the fundamental cells (triangles, squares, hexagons and cubes respectively) of the point lattices are regular, having equal sides or edges and equal angles. The number of distinct lattices (not necessarily fully regular or symmetric) that can allow displacements, rotations and reflections is small. This paper is an exploration of such a lattice: the 4-colour rhombohedral lattice.

Details of possible application areas in science and engineering are left for treatment elsewhere, but some highly speculative possibilities have already been outlined by the author [4].

## ■ The Cuboctahedral or Bell-Fuller Lattice

In the twentieth century the cuboctahedral lattice attracted the attention of Alexander Graham Bell [5] and of Richard Buckminster Fuller [6], largely because of its symmetry: each point is surrounded by twelve nearest neighbours, Figure 1, that form a quasi-regular polyhedron, the cuboctahedron, Figure 2. (In the field of crystallography a sphere-packing arrangement of cuboctahedral form is described as the cubic close-packed lattice or the face-centred cubic lattice.)

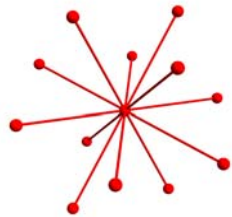

**Figure 1.** The central joint with twelve equally distributed spokes is a gusset. Alexander Graham Bell used gussets of this type to build structures, including many kites. Replication of the gusset and the links between gussets gives rise to cuboctahedral trusses, structures and space frames, which in engineering terms can be strong, rigid and light.

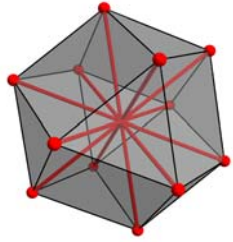

**Figure 2.** Illustration of how the Bell gusset forms a cuboctahedron

Figure 3 illustrates a size-4 Bell-Fuller lattice. The first *size* would be the trivial case of the single point at the centre of the lattice. The central point is surrounded by three more cuboctahedral *shells*. Every point in the lattice of Figure 3 belongs to one of the four shells (where the central point is regarded as shell 1).

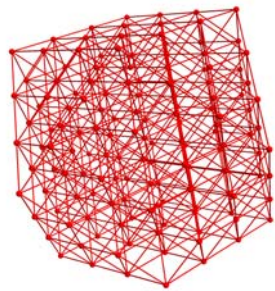

**Figure 3.** The cuboctahedral lattice is made up of cells that are either regular tetrahedra or regular octahedra. Richard Buckminster Fuller explored the geometry of this lattice.

The composite rhombohedral lattice described in this paper is perhaps even more fascinating than the cuboctahedral, or Bell-Fuller lattice: it comprises four such lattices, arranged symmetrically as in Figure 4 and Figure 5.

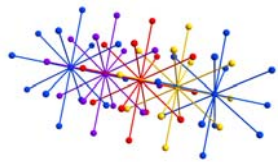

**Figure 4.** Five Bell type gussets (blue is repeated)

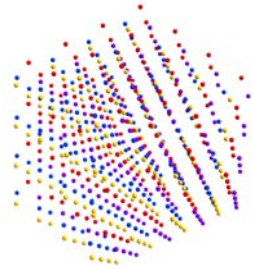

**Figure 5.** The points of four interlaced Bell-Fuller lattices

## ■ The 4-colour Rhombohedral Lattice

### ☐ Structure of the 4-colour Rhombohedral Lattice

The points of four Bell-Fuller lattices, when interlaced as shown in Figure 4 and Figure 5, comprise what is described here as a 4-colour rhombohedral lattice. All pairs of points that are nearest neighbours can be joined by links, as shown in Figure 6. The thicker gray links are described as primary spokes or axes. The black links that are opposite the primary spokes, and collinear with them, are described as secondary spokes.

The external form of the lattice shown in Figure 6 is the rhombic dodecahedron. There are twelve rhombic faces, twenty four edges and fourteen vertices. The eight vertices that correspond to the primary and secondary spokes each connect three edges of the rhombic dodecahedron and the remaining six connect four edges.

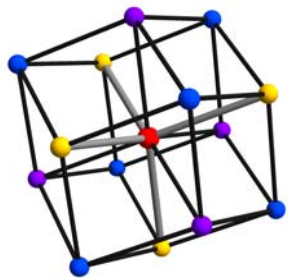

**Figure 6.** Size 2 of the 4-colour rhombohedral lattice

Figure 7 illustrates a 4-colour rhombohedral lattice of size 3. As for Figure 6, the external shape is the rhombic dodecahedron.

The term 'strut' is used here for a link of unit length between successive shells (and therefore not on any one shell) that is parallel to one of the spokes, but is not radial with respect to the centre of the overall polyhedron. Primary struts are parallel to primary spokes and secondary struts are parallel to secondary spokes.

The secondary spokes and struts have been omitted from Figure 7 to emphasise the rhombohedral nature of the cells of the lattice. When this is done there are four different orientations of the rhombohedral cells, corresponding to the orientations of the four quadrants of the rhombic dodecahedron that is shown. Each of the four quadrants of the rhombic dodecahedron is itself a rhombohedron.

The first size shell or lattice would be the trivial case of the single red point at the centre of the rhombohedral lattice. The central point is surrounded by two more dodecahedral shells. Every point in the lattice of Figure 7 belongs to one of the three dodecahedral shells (where the central point is regarded as shell 1).

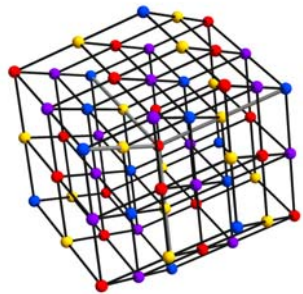

**Figure 7.** Size 3 of the 4-colour rhombohedral lattice. The secondary spokes and struts have been omitted to emphasise the rhombohedral cells of the lattice, which are oriented in accordance with the primary spokes.

## ☐ A Natural Coordinate System for the 4-colour Rhombohedral Lattice

Every point position on the 4-colour rhombohedral lattice can be described by an integer 4-tuple, as in Figure 8 and Figure 9. The full range of the set of integers can be used arbitrarily within the 4-tuple to describe the position of any point with respect to a specified 'origin' and therefore the representation is not unique. However, any such 4-tuple can be converted to zero-based form without negative integers by the addition of a unique integer to all elements that makes at least one of the elements zero and all elements non-negative. A 'standard form' 4-tuple can also likewise be defined in which at least one of the elements is unity and none are zero or negative. As negative ordinates are not used, it is considered more natural to count from the integer 1, with the origin represented as (1, 1, 1, 1), as in Figure 8. Conversions between zero-based and standard form 4-tuples are easily implemented, as explained in the following subsections.

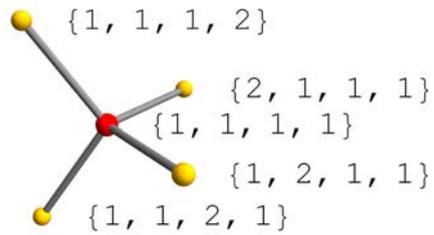

**Figure 8.** Integer coordinate system for the 4-colour rhombohedral lattice

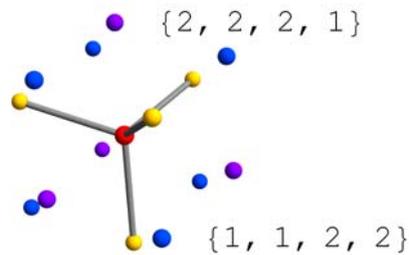

**Figure 9.** Sample integer coordinates for the 4-colour rhombohedral lattice

### *Converting 4-tuple Coordinates to Zero-based Form*

Listing 1 presents a function, `QZeroForm`, written in *Mathematica*, that converts a 4-tuple coordinate list, or vector, to zero-based form. Another version takes four separate arguments for the ordinates of the 4-tuple. Variants have also been written to deal with lists of 4-tuples or, for instance, lists of 4-tuple pairs.

```
QZeroForm[{q1_, q2_, q3_, q4_}] := Module[{qzf}, qmin = Min[q1,
q2, q3, q4];
qzf = {q1, q2, q3, q4} - {qmin, qmin, qmin, qmin}]
```

**Listing 1.** A function, written in *Mathematica*, to convert a 4-tuple to zero-based form

An example of a conversion from standard form to zero-based form in a *Mathematica* session is as follows:

**QZeroForm[{1, 2, 3, 4}]**

{0, 1, 2, 3}

### *Converting 4-tuple Coordinates to Standard Form*

Function `QStdForm`, as presented in Listing 2, converts a 4-tuple coordinate list, or vector, to standard form:

```
QStdForm[{q1_, q2_, q3_, q4_}] := Module[{qsf}, qsf =
QZeroForm[{q1, q2, q3, q4}] + {1, 1, 1, 1}]
```

**Listing 2.** A function to convert a 4-tuple to standard form

An example of a conversion from zero-based form to standard form is as follows

**QStdForm[{0, 1, 2, 3}]**

{1, 2, 3, 4}

## ☐ The Colour of a Lattice Point

Given the colour number of the origin, *nco* (an integer from 1 to 4 corresponding to the colours red, yellow, blue and violet) and a 4-tuple coordinate list ($q_1$, $q_2$, $q_3$, $q_4$) of a point, the function `CoQ` determines the colour number of the point. This is shown in Listing 3. This determination amounts to summing the four ordinates and the colour number of the origin using clock arithmetic, base 4.

```
CoQ[nco_, {q1_, q2_, q3_, q4_}] := Module[{c}, c = Mod[(q1 + q2
+ q3 + q4 + nco - 1), 4] + 1]
```

**Listing 3.** Function to give the colour number of a point on the rhombohedral lattice

For example, if the origin is red, the point (1, 2, 12, 8) is violet:

```
CoQ[1, {1, 2, 12, 8}]
4
```

## ☐ Sum, Point Inversion and Difference

### *4-tuple Vector Addition*
The sum of a pair of 4-tuples of the 4-colour rhombohedral lattice is obtained by adding the corresponding ordinates. The result would normally be converted to standard form.

### *Point Inversion and Vector Difference*
Listing 4 illustrates the implementation of a function, `QInvert`, that inverts a 4-tuple and returns the result in standard form. This is the point inversion of the 4-tuple through the origin. Each ordinate is inverted by adding one to the sum of the other three ordinates. The resulting 4-tuple is then converted to standard form.

---

```
QInvert[{q1_, q2_, q3_, q4_}] := Module[ {p1, p2, p3, p4,
qinv}, p1 = 1 + q2 + q3 + q4; p2 = q1 + 1 + q3 + q4; p3 = q1 +
q2 + 1 + q4; p4 = q1 + q2 + q3 + 1; qinv = QStdForm[{p1, p2,
p3, p4}]]
```
---

**Listing 4.** A function to obtain the additive inverse, or point inversion through the origin, of a 4-tuple

The following is an example of 4-tuple inversion as part of a *Mathematica* session:

```
QInvert[{2, 1, 11, 6}]
{10, 11, 1, 6}
```

The vector difference of two 4-tuples, $q_1 - q_2$, can be implemented by summing the first 4-tuple and the additive inverse of the second, as represented in equation (1), where I is the function `QInvert`. The result would be brought to standard form.

$$q_1 - q_2 = q_1 + \mathrm{I}(q_2) \tag{1}$$

## ☐ Length or Distance Calculations

Any planes of the lattice of Figure 6 that are made up of rhombs of unit edge length provide an invariant metric ground form [7] that allows the length of a vector in the plane to be defined. This concept can be extended to the four-tuple discrete space of the lattice. The application and implementation of this type of approach (including conversions between 4-tuple and Cartesian coordinates) has been described by Hasslberger [8] and Urner, [9] and [10], in the latter two cases using the term 'quadrays' for the four basis vectors.

The *length* or *distance* formula for a vector ($q_1$, $q_2$, $q_3$, $q_4$), which can be derived using the cosine rule, is given by equation (2).

$$|r| = \sqrt{\left[(q_1^2 + q_2^2 + q_3^2 + q_4^2) - (2/3)(q_1 q_2 + q_2 q_3 + q_3 q_4 + q_4 q_1 + q_3 q_1 + q_2 q_4)\right]}. \quad (2)$$

Listing 5 presents the function `Dist4Axis` that calculates distance, which is equivalent to calculating the length of a vector. The input is a single 4-tuple. To find the distance between two points, the input would be the vector difference of the 4-tuples of the points.

```
Dist4Axis[{q1_, q2_, q3_, q4_}] := √((q1^2 + q2^2 + q3^2 +
q4^2) - (q1 q2 + q2 q3 + q3 q4 + q4 q1 + q3 q1 + q2 q4) 2/3 )
```

**Listing 5.** Function to calculate distance on the 4-colour rhombohedral lattice

As can be seen below, this can yield a value that is the square root of a fraction that has a three in the denominator.

**Dist4Axis[{3, 1, 12, 7}]**

$$\sqrt{\frac{283}{3}}$$

An integer-only distance measure, which is equal to three-times the square of the distance between any two points on the entire lattice, is readily defined, as in equation (3).

$$s = 3\,r^2 = 3(q_1^2 + q_2^2 + q_3^2 + q_4^2) - 2(q_1 q_2 + q_2 q_3 + q_3 q_4 + q_4 q_1 + q_3 q_1 + q_2 q_4). \quad (3)$$

◻ **Scaling and Scalar Division**

*Scaling*
A 4-tuple of the 4-colour rhombohedral lattice can be scaled by multiplying the ordinates of the zero-based form by an integer. By this means the scaling is applied to the ordinate intervals. The result would be brought to standard form.

*Scalar Division*
Given two arbitrary points on the 4-colour rhombohedral lattice, the vector difference is a vector from the first to the second point. Some vectors cannot be divided by any integer other than 1: their scalar divisibility is unity. Function `QSDiv`, as presented in Listing 6, can be used to calculate the maximum number of equal (symmetric) intervals into which the vector can be divided by an integer. This is a matter of finding the greatest common divisor of the intervals of the ordinates.

```
QSDiv[{q1_, q2_, q3_, q4_}] := Module[{qsd, v}, v =
QZeroForm[{q1, q2, q3, q4}];
qsd = GCD[v[[1]], v[[2]], v[[3]], v[[4]] ]]
```

**Listing 6.** A function to find the scalar divisibility of a vector

As an example,

> **QSDiv[{3, 1, 5, 9}]**
>
> 2

## ■ Relationship Between the Discrete Rhombohedral Lattice and a Discrete Cartesian Lattice

### ☐ Nucleated Cubic (or Cartesian) Lattices Within the 4-colour Rhombohedral Lattice

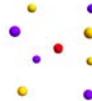

**Figure 10.** The eight vertices of the smallest cube on the 4-colour rhombohedral lattice, shown surrounding the origin of the rhombohedral lattice

The smallest cube on the 4-colour rhombohedral lattice (not counting the case of a single point) is formed by the pair of smallest regular tetrahedra on the lattice, Figure 10. This cube consists of eight points (four yellow and four violet) with an edge length of $2/\sqrt{3}$ rhombohedral lattice units. It is on shell 2 of the rhombohedral lattice. The third and fourth shells of the rhombohedral lattice also contain corresponding sets of eight points in a cubic arrangement. However, the first rhombohedral lattice size that contains a full nucleated cubic lattice is the fifth, as shown in Figure 11. (A lattice is nucleated if it has one of its points at its centre). What is remarkable is that all points of the nucleated cubic lattice have the same colour, red in this case.

The nucleated cubic lattice represented in Figure 11 is a size-2 nucleated cubic lattice. The vertices of its bounding cube are contained in the fifth (rhombic dodecahedral) shell of the 4-colour rhombohedral lattice. The spokes of the size-2 nucleated cubic lattice each contain two red points. The length of each unit link or spoke of the cubic lattice is equivalent to $4/\sqrt{3}$ rhombohedral lattice units. This red nucleated cubic lattice has all the attributes of a discrete Cartesian lattice, so this descriptor can also be used for it.

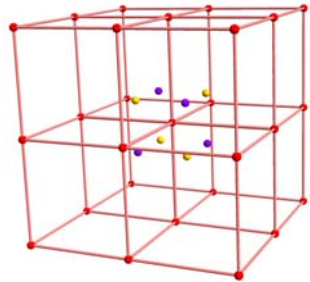

**Figure 11.** The size-2 nucleated cubic lattice (the vertices of the smallest possible cube on the 4-colour rhombohedral lattice are also shown)

A red nucleated cubic lattice of any arbitrary size can be constructed about any red point of the 4-colour rhombohedral lattice. Yellow, blue and violet cubic lattices can be constructed about any point of the respective colour, or can be produced by displacing the (originally) red cubic lattice within the 4-colour rhombohedral lattice, as was done in generating Figure 12. Whereas links of four nucleated cubic lattices are shown in Figure 12, sixteen nucleated cubic lattices (4 of each colour) are required in order to include all points of the 4-colour rhombohedral lattice. Each of the sixteen nucleated cubic lattices is distinct.

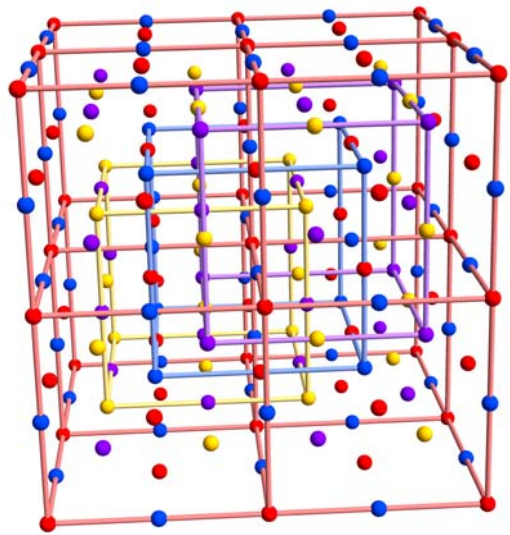

**Figure 12.** The points of the 4-colour rhombohedral lattice, with the links of four interlaced cubic lattices. Only complete lattice links are shown in this diagram. (Pastel colours are used for the links of nucleated cubic lattices.)

If all nucleated cubic lattices were shown in a diagram such as Figure 12, red links would overlap with blue links and yellow links would overlap with violet links. In combination the sixteen nucleated cubic lattices form a composite lattice where the spacing between parallel lines of links is $2/\sqrt{3}$ rhombohedral lattice units. The overlapping links have been replaced by links of this size in Figure 13 and Figure 14.

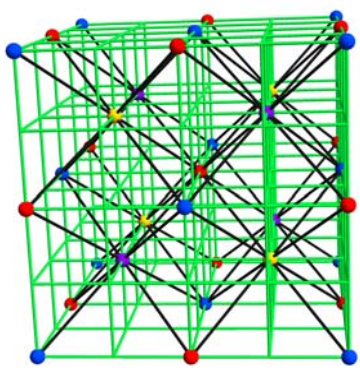

**Figure 13.** A Cartesian lattice with a diagonal length of four rhombohedral lattice units and a spacing of $2/\sqrt{3}$ rhombohedral lattice units

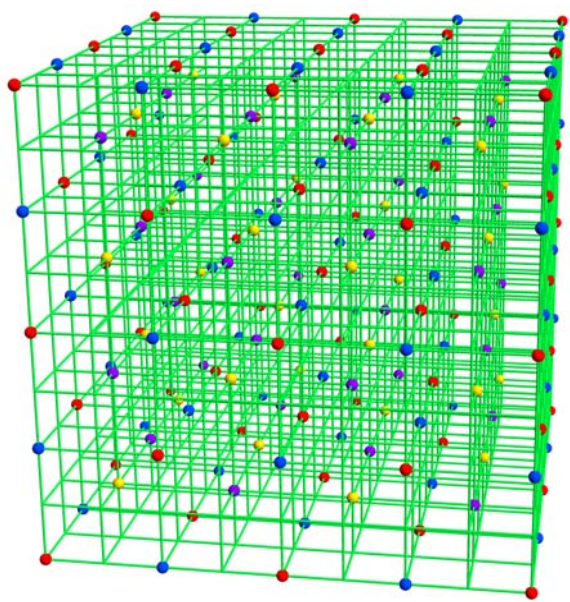

**Figure 14.** A Cartesian lattice with a diagonal length of eight rhombohedral lattice units and a spacing of $2/\sqrt{3}$ rhombohedral lattice units. The points of the 4-colour rhombohedral lattice are also shown, but without the rhombohedral lattice links.

The methodology used for conversions between discrete 4-tuple and discrete Cartesian coordinates is explained in the following subsections. The relationships used have been derived from the trigonometry of the lattices.

### *Converting 4-tuple Coordinates to Discrete Cartesian Coordinates*

Function `QtoU`, as presented in Listing 7, converts 4-tuple coordinates to Cartesian coordinates:

```
QtoU[{q1_, q2_, q3_, q4_}] := Module[{u}, u = {(q1 - q2 - q3 + q4), (q2 - q4 - q3 + q1), (q3 - q2 - q4 + q1)}]
```

**Listing 7.** Function to convert 4-tuple coordinates to Cartesian coordinates

The following is an example of applying `QtoU`:

    **QtoU[{3, 1, 12, 7}]**

    {-3, -15, 7}

### *Converting Discrete Cartesian Coordinates to 4-tuple Coordinates*

Function `UtoQ`, which is described in Listing 8, converts the Cartesian coordinate list ($u_1$, $u_2$, $u_3$) to a 4-tuple coordinate list ($q_1$, $q_2$, $q_3$, $q_4$). In each coordinate system the unit metric for that coordinate system is used. This allows the ordinates to be integers in both systems. In the Cartesian coordinate list the integers can be positive, negative or zero. The 4-tuple coordinate list that is output by the function is automatically provided in standard form, where all ordinates are positive.

```
UtoQ[{u1_, u2_, u3_}] := Module[{q}, q = {(u1 + u2 + u3)/4, (u2 - u1 - u3)/4, (u3 - u2 - u1)/4, (u1 - u2 - u3)/4}; q = q - Min[q] + {1, 1, 1, 1}]
```

**Listing 8.** Function to convert Cartesian coordinates to 4-tuple coordinates

An example of applying the function `UtoQ` is as follows:

    **UtoQ[{-3, -15, 7}]**

    {3, 1, 12, 7}

However, another example below appears, at first, to give an incorrect output because a discrete point on the Cartesian lattice is mapped to a non-existent point on the 4-colour rhombohedral lattice.

    **UtoQ[{-3, -15, 8}]**

    $\{\frac{7}{2}, 1, \frac{25}{2}, 7\}$

Every point on the discrete rhombohedral (or 4-tuple) lattice can be mapped (as above) to a point on a conventional Cartesian 3-tuple lattice with integer ordinates (positive and negative) where the base distance is $2/\sqrt{3}$ rhombohedral units. However, going from discrete Cartesian to 4-tuple coordinates, non-integer rational ordinates with a denominator of 2 would be produced wherever the Cartesian coordinate list contains a mixture of odd and even integers.

If the 4-colour rhombohedral lattice is the fundamental lattice that has to be mapped to a single discrete Cartesian coordinate system then each Cartesian coordinate list can contain only odd or only even integers: the 4-tuple lattice thus maps to Cartesian integer coordinates where the integers in each coordinate list (3-tuple) are either even or odd, but not a mixture of both. A unit of this single Cartesian lattice corresponds to $2/\sqrt{3}$ rhombohedral lattice units. This is equivalent to having sixteen interlaced Cartesian lattices, as shown in Figure 12 (where only links of four of the lattices are shown), where the length of the unit Cartesian lattice links in each lattice (red, yellow, blue or violet; four of each) is $4/\sqrt{3}$ rhombohedral lattice units.

## ■ Conclusion

The arithmetic associated with the 4-colour rhombohedral lattice is remarkably simple and possesses beautiful symmetry. Having undertaken the exploration thus far there are many enticements for the present author to delve even further.

The most important conclusion from this paper is that one 4-colour rhombohedral 4-tuple lattice maps onto sixteen distinct, discrete Cartesian lattices that each have a cell edge length that is twice the length of the smallest possible cube on the 4-colour rhombohedral lattice. The sixteen Cartesian lattices can be assimilated into a single Cartesian lattice with half the cell edge length (i.e. the edge length would be equal to the edge length of the smallest possible cube on the 4-colour rhombohedral lattice). A consequence of doing this is that some Cartesian coordinates on the single Cartesian lattice will not exist on the underlying 4-colour rhombohedral lattice.

## ■ References

## About the Author

Professor Jim McGovern is the Head of the School of Mechanical and Transport Engineering at the Dublin Institute of Technology. His research interests include mechanical engineering; applied thermodynamics; zero-emissions solutions for power, combustion and transport; transport engineering; engineering simulation and modelling; and fundamental concepts underlying all of these areas: particularly symmetry.